\documentclass[prx, twocolumn,amsmath,amssymb, superscriptaddress,longbibliography, nofootinbib]{revtex4-2}

\setcitestyle{super}
\usepackage{amsfonts}
\usepackage{amsmath}
\usepackage{amssymb}
\usepackage{array}   % table making
\usepackage{bm}      % bold math
\usepackage{braket}
\usepackage{xcolor}
\usepackage{dcolumn} % Align table columns on decimal point
\usepackage[T1]{fontenc}
\usepackage{graphicx}
\usepackage[utf8]{inputenc}
\usepackage{mathptmx}
\usepackage{tabularx}
\usepackage{url}
\usepackage{xr-hyper}
\usepackage[bookmarks=false,colorlinks,citecolor=blue]{hyperref}

\usepackage{amsthm,enumitem}
\usepackage{caption}
\usepackage{subcaption}

\makeatletter
\newcommand*{\addFileDependency}[1]{
  \typeout{(#1)}
  \@addtofilelist{#1}
  \IfFileExists{#1}{}{\typeout{No file #1.}}
}
\makeatother

\newcommand*{\myexternaldocument}[1]{
    \externaldocument{#1}
    \addFileDependency{#1.tex}
    \addFileDependency{#1.aux}
}
\myexternaldocument{supplement}

\newcommand{\makered}[1]{{\color{red}{#1}}}

\begin{document}

% ================= MAIN PAPER =================

%\renewcommand{\thetable}{\arabic{table}}
\setlength{\tabcolsep}{5pt} % for the horizontal padding
\renewcommand{\arraystretch}{1.1}

\title{Towards Quantum Computation of Bulk FeO}

\author{Daniel Staros}
\affiliation{Theoretical Division, Los Alamos National Laboratory, Los Alamos, New Mexico 87545, USA}

\author{Federico Petrovich}
\affiliation{Information Sciences, Los Alamos National Laboratory, Los Alamos, New Mexico 87545, USA}

\author{Ammar Kirmani}
\affiliation{QC Ware, Palo Alto, California 94306, USA}

\author{Daniel N. Blaschke}
\affiliation{X-Computational Physics Division, Los Alamos National Laboratory, Los Alamos, New Mexico 87545, USA}

\author{M. Cerezo}
\affiliation{Information Sciences, Los Alamos National Laboratory, Los Alamos, New Mexico 87545, USA}

\author{Jian-Xin Zhu$^*$}
\affiliation{Theoretical Division, Los Alamos National Laboratory, Los Alamos, New Mexico 87545, USA}
\affiliation{Center for Integrated Nanotechnology, Los Alamos National Laboratory, Los Alamos, New Mexico 87545}
\email{jxzhu@lanl.gov}

\date{\today}

\begin{abstract}
Quantum computation offers a promising path towards efficient calculations of the electronic structure of correlated materials. The properties of iron-based materials, which are key for advancing geological and astrophysical understanding, are notoriously difficult to calculate with classical computers. However, despite the many recent advances in quantum computing on near intermediate scale quantum hardware, a full treatment of the electronic structure of high-$Z$ atoms or molecules remains challenging. Here, we investigate the ground-state properties of molecular FeO and Fe$_2$O$_2$ model Hamiltonians using classical exact diagonalization, Variational Quantum Eigensolver (VQE), and Sample-based Quantum Diagonalization (SQD). The experimental quintet ground state of FeO is recovered classically and with quantum algorithms, while a nonet ground state is recovered for Fe$_2$O$_2$. Although raw VQE does not achieve chemical accuracy for the ground state of FeO or Fe$_2$O$_2$, SQD with two connectivity shells recovers the exact ground energy for both molecules utilizing remarkably few quantum computational resources. Overall, our results indicate that \textit{ab initio}-informed low-energy models of correlated materials, combined with SQD, offer a new alternative for scaling up calculations of ground-state properties of strongly correlated solids on existing quantum hardware.
\end{abstract}

%\pacs{}
\maketitle
%\pagenumbering{arabic}

\section{Introduction}
\label{sec:intro}
% \textit{Introduction.}
Understanding the electronic structure of iron oxides is essential for advancing research in planetary core geophysics and stellar spectroscopy.~\cite{Mazin_am1998,Cheung_jms1982} Bulk iron monoxide (FeO) is a prototypical strongly correlated system and has long served as a benchmark for developing electronic structure methods.~\cite{Fang_prl1998, Fang_prb1999, Mitas_prl2008, Dong_jpcc2023} In its solid state, FeO is a Mott insulator whose insulating behavior is incorrectly predicted to be metallic by conventional density functional approximations, illustrating the central role of electron correlation. Although advanced many-body methods can accurately describe its thermodynamic ground-state properties,~\cite{Mitas_prl2008} it remains a challenge to predict its magnetism,~\cite{Leonov_prb2015} excited states,~\cite{Rodl_prb2009,Shorikov_prb2010,Rodl_prb2012} and behavior under extreme conditions, motivating the continued development of improved correlated approaches.\cite{Leonov_prb2020,Ho_natcomm2024,Abdallah_prb2024}

Quantum computing offers a promising avenue for improving the scaling of accurate electronic structure calculations in correlated materials.\cite{Kim_sa2024, Bidart_arpc2026, robledo2025chemistry, Castaldo_arxiv2026} Near Intermediate Scale Quantum (NISQ) computation of material properties is feasible on current quantum hardware, but it is anticipated that fault tolerant quantum computation will one day facilitate electronic structure calculations that are beyond the reach of classical high-performance computers.\cite{Anurag_jcc2026, Cesario_pra2025} Until then, significant effort is still being devoted to the application of existing NISQ hardware to solve correlated electronic structure problems of both solids and molecules. Arguably more progress has been made in the case of molecules than solids due to the unfavorable scaling of the solid-state correlated electronic structure problem. Molecular successes include the quantum computation of the H$_2$ ground state energy to within chemical accuracy,\cite{Kim_sa2024} the calculation of dissociation curves for H$_2$, LiH, H$_2$O, and BeH$_2$, \cite{Zhao_npj2023, Dobrautz_arxiv2024} and other examples as detailed in a recent review article.\cite{Bidart_arpc2026} In addition, the relatively recent Sample-based Quantum Diagonalization (SQD) method offers a way to reduce the noise of NISQ energy estimates via a post-processing technique based on particle-number-conserving configuration recovery.\cite{robledo2025chemistry} In contrast, the quantum computation of solid-state properties requires more drastic dimensionality reductions of the corresponding Hamiltonians. Recent examples of quantum computational successes for solids include the coupled-cluster ansatz-based calculation of the ground state energy of BCC and FCC Fe crystals,\cite{Yamamoto_prr2022} quantum-enhanced dynamical mean field theory calculation of the cuprate Mott insulator Ca$_2$CuO$_2$Cl$_2$\cite{Selisko_npjcompmat2025} and quantum simulations of the magnetic excitation spectrum of KCuF$_3$.\cite{Lee_arxiv2026}

However, solid-state demonstrations to-date largely remain limited by workflows that restrict the quantum computation to small effective problems, such as local impurity models or simplified spin Hamiltonians, rather than realistic multi-orbital electronic Hamiltonians. These limitations motivate the development of material-specific low-energy models that retain the essential electronic structure while remaining tractable on near-term quantum hardware. Although realizations of such low-energy models on NISQ hardware exist, they are limited to the case of weak electron correlation.\cite{Sherbert_rsc2021, Sherbert_sr2022, Khandelwal_sr2025} Density Functional Theory (DFT)-informed correlated tight-binding Hamiltonians provide a natural framework for bridging this gap by enabling realistic quantum simulations beyond local impurity descriptions at substantially reduced computational cost. Further, approaches that combine DFT-informed reduction in the orbital active space with Hamiltonian symmetry reduction could allow for the allocation of valuable quantum computing resources to the \textbf{k}-point sampling that hinders the desired scaling of quantum computation in correlated solids.\cite{Khandelwal_sr2025, Yamamoto_prr2022}

Due to their classical simulability and highly correlated nature, FeO molecules form a fitting benchmark for testing quantum computation based on DFT-informed, correlated model Hamiltonians. For example, theoretical calculations of the electronic structure of the diatomic FeO molecule have had a mixed range of success, recovering its quintet ground state magnetism while failing to reproduce its experimental optical properties.\cite{Sakellaris_jcp2011} The experimental value of the first excitation energy is $0.5$ eV (4000 cm$^{-1}$),\cite{Drechsler_jcp1997} but this is significantly undershot by high-accuracy quantum chemistry calculations which predict an excitation of 300 cm$^{-1}$ (0.04 eV).\cite{Sakellaris_jcp2011} Thus, in addition to providing a first step towards quantum computation of bulk FeO, a low-energy-model-based study can provide additional insight into the role of Coulomb repulsion on the magnetic composition and multi-reference nature of the FeO molecule. Simultaneously, they can also outline whether techniques such as Variational Quantum Eigensolver (VQE) and/or SQD are sufficient to recover correlated electronic ground states. Furthermore, an idea of the computational expense scaling of a $d$-$p$ model as a function of modes can be obtained by considering the performance of VQE and SQD on molecules with variable number of FeO units, like rhombic Fe$_2$O$_2$, where the number of orbitals (even within a multi-band, low-energy model) might challenge current NISQ capabilities.

\begin{figure}[!t]
\centering
\includegraphics[width=0.48\textwidth]{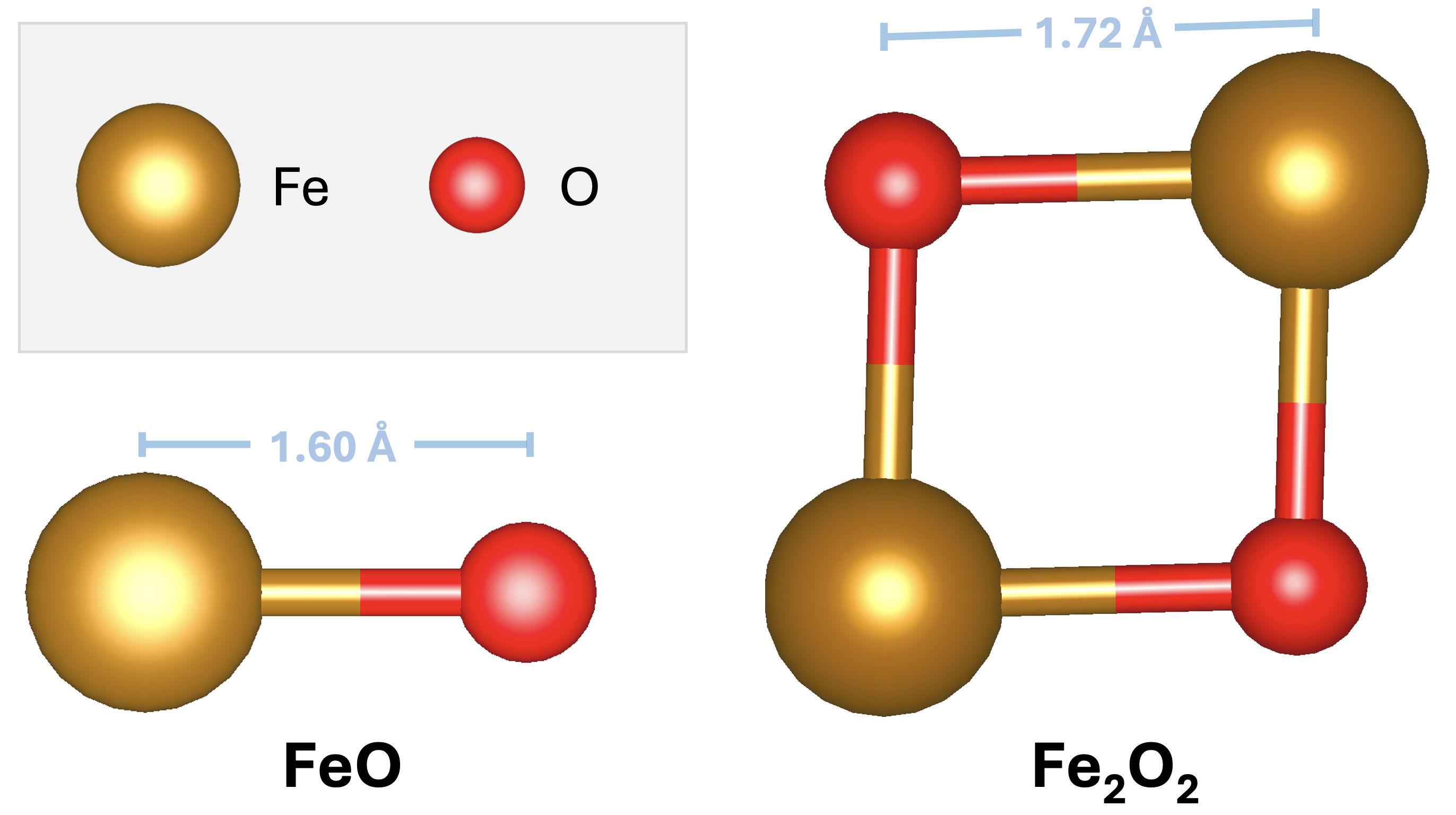}
\caption{
\textbf{Molecular geometries.} \textbf{(left)} FeO and \textbf{(right)} Fe$_2$O$_2$ molecular structures optimized via non-spin-polarized Density Functional Theory.
}
\label{fig:molecule_schematic}
\end{figure}

Here, we construct correlated effective Hamiltonians for the FeO and Fe$_2$O$_2$ molecules, for which we (i) evaluate the electronic properties via classical exact diagonalization (ED), and (ii) determine the ground state energies via both VQE and SQD calculations. First, we exactly diagonalize $d$-$p$ models of both molecules to calculate their magnetism, ground state degeneracy/multireference character, and first electronic excitation energies. Our models recover the experimental magnetism and multireference character of the FeO ground state and predict a nondegenerate magnetic ground state for Fe$_2$O$_2$.\cite{Cheung_jms1982} Then, we implement the corresponding $d$-$p$ models within the Local Unitary Cluster Jastrow (LUCJ)\cite{Shee_rsc2023} ansatz on IBM Miami to calculate ground-state energies via VQE and SQD, obtaining accurate SQD ground state energies even when VQE does not achieve chemical accuracy. Further, our approach uses remarkably few quantum computational resources, suggesting that DFT-informed low-energy models of correlated materials may provide an appealing path toward quantum computation of bulk FeO.

\section{Methods}
\label{sec:methods}
%\textit{Methods.} 

\subsection{Density Functional Theory}
\label{sec:methods_dft}
All DFT calculations are performed with Quantum ESPRESSO v7.3.\cite{Giannozzi_jpcm2009, Giannozzi_jpcm2017} Projected Wannier Functions (PWFs) and hopping amplitudes are obtained with Wannier90.\cite{Mostofi_cpc2014} We solve the non-spin-polarized Kohn Sham equations using the Perdew-Burke-Ernzerhof (PBE) approximation of the exchange correlation potential.\cite{Perdew_prl1996} The core electrons are treated with high quality correlation-consistent Effective Core Pseudopotentials (ccECP).\cite{kincaidCorrelationConsistentEffective2022, wang_new_2022} A cutoff energy of 320 Ry is used for the plane-wave basis set. The DFT and PWF calculations are performed at the $\Gamma$-point unless otherwise specified. Band structure calculations of the FeO molecule performed along the $\Gamma$-X-M-$\Gamma$ high symmetry path verify the absence of inter-image interactions (see the Supplementary Information), indicating that a $15 \times 15 \times 15$ Å$^3$ box is sufficient for our analysis. 

For each molecule, the charge density is self-consistently converged, energy eigenvalues are obtained via a non-self-consistent field calculation, and the orbital character of the Kohn-Sham solutions is determined by projection onto atomic wavefunctions. This determines the bands with dominant Fe $3d$ and O $2p$ character; the $d$-like and $p$-like PWFs and their corresponding hopping amplitudes are obtained from these bands.

\subsection{Low-energy models}
\label{sec:methods_dp}
The valence electronic structure of the FeO and Fe$_2$O$_2$ molecules nominally consists of twelve electrons occupying sixteen spin orbitals per FeO unit; there are ten 3$d$ spin orbitals on each Fe atom, and six $2p$ spin orbitals on each O atom. We work in the fermionic Hilbert space defined by this description of the valence structure.  We denote by $\hat p_{i,s}^{\dagger}$ ($\hat p_{i,s}$) the creation (annihilation) operator for an electron with spin $s$ in the $i^{\textnormal{th}}$ O $p$ orbital, and by
$\hat d_{j,s}^{\dagger}$ ($\hat d_{j,s}$) the corresponding operators for the $j^{\textnormal{th}}$ Fe $d$ orbital. In addition, we will call
\begin{subequations}
\begin{eqnarray}
\hat{G}^{p,p}_{i,i'}(s) &=& \tfrac{1}{2} (\hat{p}_{s,i}^\dag \hat{p}_{s,i'}+\hat{p}^\dag_{s,i'} \hat{p}_{s,i}), \\
\hat{G}^{d,d}_{j,j'}(s) &=& \tfrac{1}{2} (\hat{d}_{s,j}^\dag \hat{d}_{s,j'}+\hat{d}^\dag_{s,j'} \hat{d}_{s,j}), \\
\hat{G}^{p,d}_{i,j}(s) &=& \tfrac{1}{2}(\hat{p}_{s,i}^\dag \hat{d}_{s,j} + \hat{d}^\dag_{s,j} \hat{p}_{s,i}).
\end{eqnarray}
\end{subequations}
The fermionic Hamiltonian is written as
\begin{equation}
\hat H=\left(\sum_{s\in\{\uparrow,\downarrow\}} \hat h^p(s)+\hat h^d(s)+\hat h^{pd}(s)\right)+\hat{V}.
\label{Ham}
\end{equation}
The one-particle contribution is separated into Fe $d$, O $p$, and $d$-$p$ hybridization components, such that we have 
\begin{subequations}
\begin{eqnarray}
\hat h^p(s) &\!\!=\!\!& \sum_{\mu=x,y,z}\epsilon_\mu^p\,\hat G^{p,p}_{\mu, \mu}(s), \\
\hat h^d(s) &\!\!=\!\!& \sum_{\mu=x,y,z}\epsilon_\mu^d\,\hat G^{d,d}_{\bar \mu,\bar \mu}(s) + \sum_{\alpha,\beta=y,z} h^d_{\tilde \alpha \tilde \beta} \hat G^{d,d}_{\tilde \alpha,\tilde \beta}(s), \\
\hat h^{pd}(s) &\!\!=\!\!& \sum_{\alpha=y,z} t^{p,d}_{x,\tilde \alpha} \hat G^{p,d}_{x, \tilde \alpha}(s) \!+\! \sum_{\alpha \neq \beta=y,z} t^{p,d}_{\alpha,\bar \beta} \hat G^{p,d}_{\alpha, \bar \beta}(s), \;\;\;\;\;\;\;
\end{eqnarray}
\label{Hone}
\end{subequations}

\noindent
where the onsite energies $\varepsilon$ and hopping amplitudes $t$ for each molecule are obtained from our DFT/PWF workflow as tabulated in the Supplementary Information. The potential acts only among the Fe $d$ orbitals and is of the form
\begin{equation}
\hat V={} U \sum_{j} \hat D_{jj}^{\uparrow \downarrow} 
+\sum_{j \neq k} U' \hat{D}_{jk}^{\uparrow \downarrow} + \tfrac{U'-J_H}{2} (\hat{D}_{jk}^{\uparrow \uparrow}+\hat{D}_{jk}^{\downarrow \downarrow})\,,
\end{equation}
where
\begin{equation}
\hat{D}^{ss'}_{jk} = \tfrac{1}{2} \left(G^{d,d}_{j,j}(s) G^{d,d}_{k,k}(s') + G^{d,d}_{k,k}(s) G^{d,d}_{j,j}(s') \right)
\end{equation}
\noindent
with intra-orbital Coulomb repulsion $U=4.5$, Hund exchange $J_H=0.9$, and inter-orbital Coulomb repulsion $U'=U-2J_H=2.7$. The value of $U$ is chosen based on previously published PBE DFT-based calculations of optimal $U$ values for the FeO$^+$ molecule.\cite{Kulik_jcp2011} Specifically, previous work obtained self-consistent field and linear response $U$ values of 5.5 eV and 6.25 eV respectively for quartet FeO$^+$ with a bond distance of 1.58 Å, which is 1\% less than the FeO bond length employed in the current work. We thus choose $U=4.5$ eV to account for the decreased orbital contraction and thus decreased effective $U$ in neutral FeO relative to charged FeO$^+$. This choice of $U$ is consistent with the PBE approximation inherent to our DFT-obtained hopping matrix. We also choose a physically reasonable value of $J_H=0.9$ eV. Although these parameters do not necessarily translate directly to the Fe$_2$O$_2$ molecule, we keep them the same to maintain consistency in our approach. This choice does not impact the key quantum computational findings of our study.

\subsection{Variational quantum state preparation}
\label{sec:methods_vqe}

To construct low-energy trial states for the FeO and Fe$_2$O$_2$ Hamiltonians, we employ VQE.\cite{peruzzo2014variational,cerezo2020variationalreview} In VQE, a parameterized quantum circuit prepares a state
$\ket{\Psi(\boldsymbol{\theta})}=U(\boldsymbol{\theta})\ket{\Phi_{\mathrm{ref}}}$,
where $\ket{\Phi_{\mathrm{ref}}}$ is an easy-to-prepare fiducial state and the circuit parameters are optimized by minimizing
\begin{equation}
E_{\mathrm{VQE}}(\boldsymbol{\theta})
=
\bra{\Psi(\boldsymbol{\theta})}
\hat H
\ket{\Psi(\boldsymbol{\theta})}.
\end{equation}
The variational principle guarantees that $E_{\mathrm{VQE}}$ is greater than or equal to the true ground state energy. On a quantum computer, the energy is estimated by decomposing $\hat H$ into efficiently measurable operators, estimating their expectation values, and classically reconstructing the energy. A classical optimizer then updates $\boldsymbol{\theta}$, and the quantum-classical loop is repeated until convergence is achieved.

One of the central challenges in VQE is the choice of the ansatz $U(\boldsymbol{\theta})$. Although highly expressive, problem-agnostic circuits explore large regions of Hilbert space, sometimes at the cost of trainability.\cite{holmes2021connecting} In particular, sufficiently unstructured or expressive circuits can exhibit barren plateaus, where the cost function or its gradients become exponentially concentrated with system size.\cite{mcclean2018barren,cerezo2020cost,larocca2024review} Resolving a useful optimization direction can then require exponentially many measurements. Problem-inspired ansatzes that incorporate physical structure directly into the variational circuit can potentially help address the tradeoff between expressibility and trainability. While such constructions can preserve relevant symmetries and restrict the variational search to physically motivated regions of Hilbert space, they are not generically free of barren plateaus.\cite{larocca2021diagnosing} Thus, problem-inspired ansatzes are typically also accompanied by informed parameter initialization which avoids generic random initializations and starts the optimization near physically meaningful states.\cite{grant2019initialization,kulshrestha2022beinit,duffield2023bayesian,heyraud2023efficient,zhang2022escaping,park2023hamiltonian,wang2023trainability,park2024hardware,shi2024avoiding,puig2024variational,kulshrestha2026exponentially}

Here, we use a one-layer LUCJ ansatz,\cite{Shee_rsc2023,robledo2025chemistry,lin2025improved}
\begin{equation}
\ket{\Psi(\boldsymbol{\theta})}
=
\hat K_{\mathrm{post}}
\hat U_J
\hat K_{\mathrm{pre}}
\ket{\Phi_{\mathrm{ref}}},
\label{eq:lucj}
\end{equation}
where $\ket{\Phi_{\mathrm{ref}}}$ is a Slater determinant. The operators
$\hat K_{\mathrm{pre}}$ and $\hat K_{\mathrm{post}}$ are
particle-number-conserving fermionic Gaussian unitaries describing orbital rotations,
\begin{equation}
\hat K_{\nu}
=
e^{\hat\kappa_{\nu}},
\quad
\hat\kappa_{\nu}
=
\sum_{p,q} \sum_{s}
\kappa^{(\nu,s)}_{pq}
\hat c^\dagger_{p,s}\hat c_{q,s},
\quad
\nu\in\{\mathrm{pre},\mathrm{post}\},
\label{eq:orbital_rotation}
\end{equation}
with anti-Hermitian $\kappa^{(\nu,\sigma)}$. These transformations preserve
the number of up- and down-spin fermions ($N_{\uparrow}$ and $N_{\downarrow}$ respectively) and map Slater determinants to Slater determinants. In the circuit, they are implemented as networks of two-mode Givens rotations. Correlations beyond the fermionic Gaussian manifold are introduced by the Jastrow correlator
\begin{equation}
\hat U_J=e^{i\hat J},
\qquad
\hat J
=
\frac{1}{2} \sum_{p,q} \; \sum_{s,s'}
J^{ss'}_{pq}
\hat n_{p,s}\hat n_{q,s'},
\label{eq:j_operator}
\end{equation}
where $\hat n_{p,s}=\hat c^\dagger_{p,s}\hat c_{p,s}$ and
$J^{ss'}_{pq}$ are real. Since $\hat U_J$ is diagonal in the occupation basis, it introduces configuration-dependent phases determined by the simultaneous occupation of pairs of spin orbitals. In the LUCJ ansatz, the nonzero couplings are restricted by the connectivity of the quantum processor, reducing the SWAP overhead associated with an unrestricted UCJ ansatz.

The three blocks in Eq.~\eqref{eq:lucj} play complementary roles. The first orbital rotation $\hat K_{\mathrm{pre}}$ generates a superposition of occupation-number configurations, the Jastrow layer introduces many-body correlations through occupation-dependent phases, and $\hat K_{\mathrm{post}}$ provides an additional variational orbital rotation after the correlated layer. A single LUCJ layer therefore provides a compromise between physical structure, expressibility, trainability, and hardware implementation cost, and has also been successfully employed as a sampler in large-scale electronic-structure calculations.\cite{robledo2025chemistry}

Finally, because the ansatz conserves $N_\uparrow$ and $N_\downarrow$ separately, the optimization occurs within a fixed spin-projection sector,
\begin{equation}
M_S=\frac{N_\uparrow-N_\downarrow}{2}.
\end{equation}
A single VQE calculation therefore cannot move between sectors with different values of $(N_\uparrow,N_\downarrow)$. For each molecule, we perform separate optimizations in the relevant particle-number sectors and retain the lowest-energy solution. Further details of the reference determinants, circuit parametrization, initialization, optimization, and mapping to the quantum processor are given in the Supplementary Information.

\subsection{Sample-based Quantum Diagonalization on IBM's quantum computer}
\label{sec:methods_sqd}

Even with a problem-inspired ansatz and informed initialization, VQE remains
a nonconvex optimization over a restricted nonlinear variational manifold. Convergence to the exact ground state is therefore not guaranteed, and the
optimization typically becomes trapped in suboptimal regions of parameter space where
the ansatz itself may not contain the exact ground state.\cite{you2021exponentially,fontana2022nontrivial,anschuetz2022beyond,anschuetz2021critical,larocca2021diagnosing,kulshrestha2026exponentially} In the present work,
we therefore use the optimized VQE state, whatever low-energy solution is
obtained, as a physically informed sampler for the subsequent
SQD calculation. Using the optimized VQE state as an informed sampler is the basic idea behind quantum-selected configuration interaction and
SQD.\cite{kanno2023quantum,robledo2025chemistry} Rather than requiring the quantum
circuit itself to represent the exact ground state, measurements of the VQE
state are used to identify a reduced many-body subspace in which the
Hamiltonian is subsequently diagonalized classically.

To obtain the samples used in SQD, we execute the optimized variational
circuits on an IBM quantum computer. The circuits are first optimized using the
noiseless VQE simulations described above, after which the optimized parameter
values are fixed and transferred directly to the hardware implementation. We
do not re-optimize the circuits on the quantum processor. Instead, the
parameter-bound LUCJ circuits are compiled on the IBM device and measured in
the computational basis to generate fermionic occupation-number samples. In the Jordan-Wigner representation used here, measuring the optimized LUCJ
state in the computational basis produces bitstrings corresponding directly
to fermionic occupation-number configurations, or equivalently Slater
determinants. If $
\ket{\Psi}
=
\sum_x c_x\ket{x}$, then a computational-basis measurement returns the determinant $\ket{x}$ with
probability $|c_x|^2$. Repeated measurements therefore preferentially identify
configurations with appreciable weight in the variational state.

Given that we run the circuit on noisy hardware, errors can produce bitstrings outside the particle-number
sector preserved by the ideal LUCJ circuit. We therefore strictly postselect
the measured configurations onto the desired
$(N_\uparrow,N_\downarrow)$ sector. We do not perform any additional configuration-recovery
procedure.\cite{robledo2025chemistry} Where applicable, we further complete the
sampled determinant set using the exact global spin-reversal symmetry of the
Hamiltonian.

Let $\mathcal S_0$ denote the set of retained determinants. The Hamiltonian is
then projected onto the subspace spanned by these configurations,
\begin{equation}
\hat H_{\mathcal S_0}
=
\hat P_{\mathcal S_0}\hat H\hat P_{\mathcal S_0},
\qquad
\hat P_{\mathcal S_0}
=
\sum_{\ket{x}\in\mathcal S_0}
\ket{x}\bra{x},
\end{equation}
and diagonalized classically. Since the Hamiltonian matrix elements
are computed classically, the resulting lowest Ritz energy remains a
variational upper bound to the exact ground-state energy.

Finally, we enlarge the sampled subspace using the connectivity of the
Hamiltonian in Fock space. Starting from a set $\mathcal S_m$, we define
\begin{equation}
\mathcal S_{m+1}
=
\mathcal S_m
\cup
\left\{
\ket{y}:
\exists\,\ket{x}\in\mathcal S_m
\ \text{such that}\
\bra{y}\hat H\ket{x}\neq 0
\right\}.
\label{eq:hamiltonian_connectivity}
\end{equation}
We compare the raw sampled space with the spaces obtained after one and two
such Hamiltonian-connectivity expansions. This allows the quantum circuit to
identify the relevant region of Fock space, while the Hamiltonian itself is
used to recover configurations that may have small sampling probability but
remain important for the low-energy eigenstates.

\section{Results}
\label{sec:results}
%\textit{Results.} 

\subsection{DFT Structure Relaxation}
Our non-spin-polarized DFT calculations result in an energetic minimum when the Fe-O bond length equals $1.60$ Å (Figure \ref{fig:molecule_schematic}); this is fortuitously just 0.3\% greater than the bond length predicted via multireference configuration interaction and coupled cluster with singles, doubles and triples.\cite{Sakellaris_jcp2011} The atomic coordinate relaxation for Fe$_2$O$_2$ results in the rhombic geometry displayed in Figure \ref{fig:molecule_schematic} with an Fe-O bond length equal to 1.72 Å, which, predictably, undershoots previously published B3LYP DFT results by $\sim4\%$.\cite{Cao_faraday1998} The optimized Fe$_2$O$_2$ geometry exhibits a point group symmetry that is consistent with previous DFT/coupled-cluster determination of the $^7B_{2u}$ ground state of the Fe$_2$O$_2$ molecule.\cite{Cao_faraday1998} These are the geometries employed for DFT/PWF calculation of the hopping amplitudes that parameterize our low-energy models.

\subsection{Classical Exact Diagonalization}
First, we present results for our classical exact diagonalization analysis of the diatomic FeO molecule. The ground-state energy ($E_{GS}$), the degeneracy of the ground-state manifold (deg$(\Psi_{GS})$), and the spin eigenvalues of the ground-state manifold, are presented in the first row of Table \ref{tab:classical_summary}. The reflection symmetries of the diatomic FeO molecule generate a $\mathbb{Z}_2 \times \mathbb{Z}_2$ subgroup of the $C_{\infty v}$ point group, allowing decomposition of the twelve-electron Hilbert space into four sectors,
\begin{equation}
    \hat{H}_{12} = \bigoplus_{\pi_y,\pi_z} \hat{H}_{12}^{(\pi_y,\pi_z)}
\end{equation}
which have dimensions dim $\hat{H}_{12}^{(+,+)}=$ dim $\hat{H}_{12}^{(-,-)} =452$ and dim $\hat{H}_{12}^{(+,-)}=$ dim $\hat{H}_{12}^{(-,+)}=448$. Restricting the Hamiltonian to the twelve-particle, positive reflection sector $\hat{H}_{12}^{(+,+)}$, exact diagonalization results in a twofold degenerate ground-state energy of $E_0=-47.76$ eV. This doubly degenerate ground-state manifold consists of oppositely polarized quintet states for the diatomic FeO molecule, which is consistent with the experimentally measured $^5 \Delta_i$ ground state of gaseous FeO.\cite{Cheung_jms1982} The spin magnetization within the ground-state manifold is characterized by the spin projection along the quantization axis,
\begin{equation}
    \hat{S}_z = \frac{1}{2} \sum_{r=1}^{8}\left(\hat{n}_{r,\uparrow}-\hat{n}_{r,\downarrow}\right) = \frac{\hat{N}_{\uparrow}-\hat{N}_{\downarrow}}{2},
\end{equation}
from which the spectrum of spin eigenvalues presented in Table \ref{tab:classical_summary} is obtained by diagonalizing $V_{\mathcal{G}}^{\dagger} \hat{S}_z V_{\mathcal{G}}$ where the isometry $\hat{V}_{\mathcal{G}} = \left(\ket{\psi_1}, \ket{\psi_{2}}\right)$ is composed of the two degenerate ground eigenstates. The first electronic excitation energy of our model is approximately 50\% that of the experimentally measured value of $\sim$ 4000 cm$^{-1}$. Somewhat fortuitously, our first electronic excitation energy is about six times larger in magnitude than first excitation energies previously obtained via configuration interaction and coupled-cluster calculations.\cite{Sakellaris_jcp2011}

\begin{table}[ht]
    \caption{Ground-state energy, degeneracy, spin eigenvalues, and first electronic excitation energy of FeO and Fe$_2$O$_2$ from classical exact diagonalization.}
    \label{tab:classical_summary}
    \centering
    \begin{tabular}{ccccc}
        \hline\hline
        Molecule & $E_{GS}$ (eV) & deg$(\Psi_{GS})$ & spec$\left(\tilde{S}_Z^{GS}\right)$ & $E^*-E_{GS}$ (cm$^{-1}$) \\
        \hline
        FeO & -47.7639 & 2 & $\left\lbrace -2,2 \right\rbrace$ & 2167.07 \\
        Fe$_2$O$_2$ & -10.8400  & 1 & 4 & 193.80 \\
        \hline\hline
    \end{tabular}
\end{table}

The $D_{2h}$ molecular point group symmetry of the Fe$_2$O$_2$ molecule results in a $24$-electron Hilbert space that decomposes as
\begin{equation}
\hat{H}_{24}
=
\bigoplus_{\Gamma\in D_{2h}}
\hat{H}_{24}^{(\Gamma)}.
\end{equation}
For the 32 spin orbitals in our Fe$_2$O$_2$ model, the dimensions of the eight $N_e=24$ sectors are provided in Table \ref{tab:irrep_dimensions_fe2o2}.
\begin{table}[ht]
    \caption{Dimensions of the symmetry-resolved irreducible representation sectors in Fe$_2$O$_2$.}
    \label{tab:irrep_dimensions_fe2o2}
    \centering
    \begin{tabular}{cccc}
        \hline\hline
        Irrep & Dimension & Irrep & Dimension \\
        \hline
        $A_g$    & 1,314,524 & $A_u$    & 1,314,960 \\
        $B_{1g}$ & 1,315,616 & $B_{1u}$ & 1,314,960 \\
        $B_{2g}$ & 1,314,960 & $B_{2u}$ & 1,314,160 \\
        $B_{3g}$ & 1,314,960 & $B_{3u}$ & 1,314,160 \\
        \hline\hline
    \end{tabular}
\end{table}
Thus the purely spatial $B_{2g}$ reduction lowers the 24-electron problem from
dimension 10,518,300 to
1,314,960. Exact diagonalization in this sector of the Hamiltonian results in a nondegenerate ground state with energy $E_{GS}=-10.84$ eV and a nonet spin configuration. The rhombic $^7B_{2u}$ ground state of Fe$_2$O$_2$ was previously established via high-accuracy quantum chemistry calculations, which also predicted a septet spin configuration for the ground state.\cite{Cao_faraday1998} The difference between the ground state predicted by our model and that observed experimentally is most likely a consequence of using $U$ and $J$ values that were not optimized from first principles for Fe$_2$O$_2$, but were chosen to ensure consistency across model definitions.

\begin{figure*}[t]
    \centering
    \includegraphics[width=.9\textwidth]{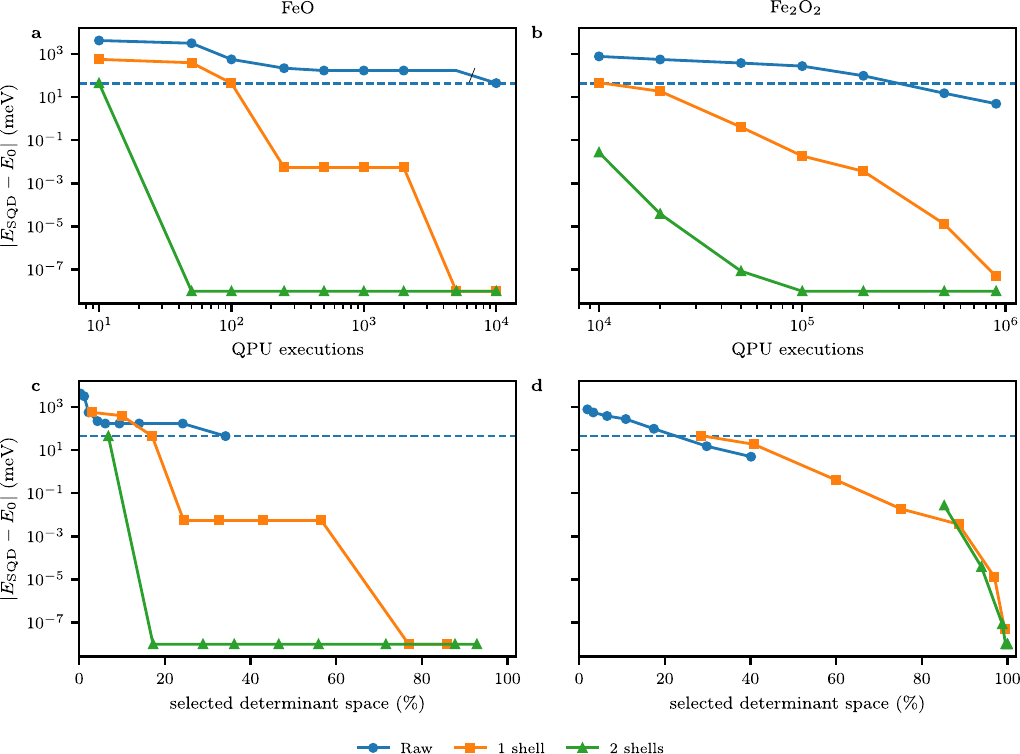}
    \caption{
    \textbf{Sample-based Quantum Diagonalization from IBM hardware measurements.}
    Ground-state energy error obtained from the raw hardware-selected determinant
    space and after one and two Hamiltonian-connectivity expansions for
    \textbf{(a,c)} FeO and \textbf{(b,d)} Fe$_2$O$_2$.
    Panels \textbf{(a,b)} show the error as a function of the number of QPU
    executions, while panels \textbf{(c,d)} show the same error as a function of
    the fraction of the relevant determinant space retained in the projected
    diagonalization. The dashed horizontal line denotes chemical accuracy,
    $1~{\rm kcal/mol}\simeq 43~{\rm meV}$.
    For FeO, Hamiltonian-connectivity expansion rapidly recovers the ground-state
    energy while retaining only a small fraction of the full determinant space.
    For Fe$_2$O$_2$, the same expansion strongly improves the energy, although the
    two-shell space approaches the full $(N_\uparrow,N_\downarrow)=(16,8)$ sector
    as the sampling budget increases.
    }
    \label{fig:vqe_sqd}
\end{figure*}

\subsection{Sample-based Quantum Diagonalization}
\label{sec:results_sqd}

Here we present the results obtained using the LUCJ circuits executed on IBM hardware as samplers for SQD.
Figure~\ref{fig:vqe_sqd} shows the resulting ground-state energy error for
FeO and Fe$_2$O$_2$ relative to the corresponding classical benchmark. The upper panels report the error as a function of the
number of hardware executions, while the lower panels show the same data as a
function of the fraction of the relevant determinant space retained in the
projected diagonalization. The dashed line indicates chemical accuracy,
$1~{\rm kcal/mol}\simeq 43~{\rm meV}$.

For FeO, diagonalization in the raw determinant support obtained directly
from hardware sampling converges relatively slowly. In contrast, adding
Hamiltonian-connectivity shells produces a dramatic improvement. A single
shell reduces the energy error by several orders of magnitude, while the
two-shell space reaches chemical accuracy already at the smallest hardware
sampling budget considered and rapidly converges to the exact energy within
numerical precision (Figure~\ref{fig:vqe_sqd}\makered{a}). Importantly, this convergence
is not simply the result of recovering the complete Hilbert space. As shown
in Figure~\ref{fig:vqe_sqd}\makered{c}, numerical-precision agreement is obtained while
retaining only a small fraction of the full FeO determinant space. The
hardware samples therefore identify a compact region of Fock space from which
Hamiltonian connectivity efficiently reconstructs the missing components of
the ground state.

The larger Fe$_2$O$_2$ problem displays the same qualitative hierarchy, but
also illustrates the limitations of this compression. The raw sampled support
requires substantially more hardware sampling to reach chemical accuracy,
whereas one connectivity shell reaches this regime much earlier. With two
shells, the error is already approximately $2.7\times10^{-2}$~meV at
$10^4$ QPU executions and decreases rapidly thereafter
(Figure~\ref{fig:vqe_sqd}\makered{b}). However, the selected-space representation in
this case is considerably less compressed: the two-shell space already
contains approximately $85\%$ of the $(N_\uparrow,N_\downarrow)=(16,8)$
sector at the smallest sampling point and approaches the complete sector as
the number of samples increases. Hence, the extremely small
two-shell errors at large sample counts should be interpreted as convergence
toward full classical diagonalization rather than as evidence for strong
Hilbert space compression.

The one-shell Fe$_2$O$_2$ results provide a more informative intermediate
regime. Chemical accuracy is reached while retaining only a fraction of the
full determinant sector, demonstrating that the Hamiltonian expansion can
substantially improve upon the raw quantum-selected support before the
selected space saturates. Taken together, the two molecules illustrate the two roles played by SQD in the present calculations. The quantum processor provides a physically informed sampling of Fock space, while Hamiltonian connectivity systematically enlarges this support and allows the subsequent classical Ritz calculation to escape the restricted nonlinear manifold of the original VQE ansatz. FeO exhibits strong compression, whereas Fe$_2$O$_2$ provides a more demanding scaling test in which increasing accuracy comes at the cost of a progressively larger classical subspace.

Finally, we note that for the Fe$_2$O$_2$ model, exact diagonalization in the
$(N_\uparrow,N_\downarrow)=(16,8)$ sector, corresponding to
$M_S=4$, gives a lowest energy 
\begin{equation}
E_0=-10.839851578034~{\rm eV}.
\end{equation}
The value above is therefore used as the exact reference energy for the
Fe$_2$O$_2$ SQD results in Fig.~\ref{fig:vqe_sqd}. More importantly, the ground state obtained via SQD calculation lies in the
$M_S=4$ sector. This nonet magnetic configuration differs from the septet magnetic configuration
previously reported for rhombic Fe$_2$O$_2$ in high-accuracy quantum-chemistry
calculations.\cite{Cao_faraday1998} The discrepancy should be understood as a difference in the
magnetic state favored given the choice of $U$ and $J$ parameters in our low-energy Hamiltonian, which are likely different than the optimal $U$ and $J$ parameters that might be obtained from additional first-principles calculations of Fe$_2$O$_2$.

\section{Conclusions}
\label{sec:conclusions}
%\textit{Conclusions.} \DS{In conclusion...} 

We have presented a new low-energy-model-based approach for scalable quantum computation of correlated FeO materials using the FeO and Fe$_2$O$_2$ molecules as benchmarks. In addition to providing an energetic benchmark for our SQD calculations, we demonstrated that classical exact diagonalization of our \textit{ab initio}-informed $d$-$p$ model of FeO recovers the experimentally measured ground state magnetism given the appropriate interaction parameters, and fortuitously predicts a first electronic excitation energy which is closer to experiment than previous quantum chemical estimates. We found that VQE does not need to achieve chemical accuracy to be a useful tool in the quantum computation of FeO and Fe$_2$O$_2$ ground-state properties. The best one-layer LUCJ state is still 0.173 eV above the ground state energy and has a large eigenstate residual, but its 99.1\% ground-space weight concentrates the samples on a compact, highly relevant determinant set. SQD then replaces the nonlinear variational restriction with exact diagonalization in a physically selected linear subspace. One connectivity shell recovers essentially all of the ground-state energy, and two shells recover the exact ground doublet with machine-precision residuals in only 184 determinants. 

The primary limitation of our framework lies in the need to perform exact diagonalization on the Hamiltonian projected into the VQE-enhanced determinantal support, which we have shown grows quickly as a function of hopping-allowed modes in the Hamiltonian. Despite this, we anticipate that the approach presented here will be almost directly applicable to solid FeO, with the exception that the periodic dependence of the interaction potential will need to be incorporated. As periodic FeO consists of repeating units of the diatomic FeO molecule, we anticipate that VQE sampling of the Fock space and subsequent SQD will result in projected Hamiltonians that can feasibly be exactly diagonalized within their \textbf{k}-point sub-blocks. The main additional cost in solid FeO will thus be the need to perform SQD at hundreds to thousands of \textbf{k}-points, which can be partly aided by Brillouin zone symmetrization.

\section{Associated Content}
\label{sec:associated content}
%\textit{Associated Content.} 
The code, input files, and output files generated in this work are available upon reasonable request. See the Supplementary Information for further details regarding the DFT calculations and hopping amplitudes used to parameterize the $d$-$p$ models, the variational circuit optimization, and the hardware implementation of SQD.

\section{AI Disclosure Statement}
\label{sec:AI}
We acknowledge that ChatGPT was used for some portions of the literature review, to improve VQE and SQD code, to polish phrasing in the third paragraph of the introduction, to automate the symmetry analysis for the FeO and Fe$_2$O$_2$ molecules, and to generate the table of Fe$_2$O$_2$ hopping amplitudes in the Supplementary Information. All ChatGPT output was verified by the authors and corrected when needed.

\section{Acknowledgments}
\label{sec:acknowledgments}
%\textit{Acknowledgments.} 
We thank Rahul Hingorani for useful and productive discussions. This work was carried out under the auspices of the U.S. Department of Energy (DOE) National Nuclear Security Administration under Contract No. 89233218CNA000001. It was supported by the LANL LDRD Program with project number 20260043DR, ``Fault-Tolerant Quantum Simulation for Fermionic Systems: From Nuclear to Condensed Matter Physics”, and was performed, in part, at the Center for Integrated Nanotechnologies, an Office of Science User Facility operated for the U.S. Department of Energy (DOE) Office of Science in partnership with the LANL Institutional Computing Program for computational resources. J.-X.Z. was supported in part by the NNSA Advanced Simulation and Computing Program. We acknowledge the use of IBM Quantum services for this work. The views expressed are those of the authors, and do not reflect the official policy or position of IBM or the IBM Quantum team.

\bibliography{main,quantum}

@article{Castaldo_arxiv2026,
  title = {{Utility-scale quantum computational chemistry}},
  journal = {arXiv},
  author = {Castaldo, D. and Reiher, M.},
  year = {2026},
  url = {https://arxiv.org/html/2603.19081v1},
}

@article{Bidart_arpc2026,
    author = {Bidart, A. and Vaish, P. and Kabengele, T. and Pang, Y. and Liu, Y. and Rubenstein, B.},
    title = {{Quantum Computing Beyond Ground-State Electronic Structure: A Review of Progress Toward Quantum Chemistry Out of the Ground State}},
    journal = {Annual Review of Physical Chemistry},
    year = {2026},
    volume = {77},
    url = {https://doi.org/10.1146/annurev-physchem-082624-084635}
}

@article{Kim_sa2024,
    author = {Kim, B. and Hu, K.-M. and Sohn, M.-H. and Kim, Y. and Kim, Y.-S. and Lee, S.-W. and Lim, H.-T.},
    title = {{Qudit-based variational quantum eigensolver using photonic orbital angular momentum states}},
    journal = {Science Advances},
    year = {2024},
    volume = {10},
    url = {https://doi.org/10.1126/sciadv.ado3472}
}

@article{Dobrautz_arxiv2024,
    author = {Dobrautz, W. and Sokolov, I. O. and Liao, K. and López Ríos, P. and Rahm, M. and Alavi, A. and Tavernelli, I.},
    title = {{Ab Initio Transcorrelated Method enabling accurate Quantum Chemistry on near-term Quantum Hardware}},
    journal = {arXiv},
    year = {2024},
    url = {https://doi.org/10.48550/arXiv.2303.02007}
}

@article{Zhao_npj2023,
    author = {Zhao, L. and Goings, J. and Shin, K. and Kyoung, W. and Fuks, J. I. and Rhee, J.-K. K. and Rhee, Y. M. and Wright, K. and Nguyen, J. and Kim, J. and Johri, S.},
    title = {{Orbital-optimized pair-correlated electron simulations on trapped-ion quantum computers}},
    journal = {npj Quantum Information},
    year = {2023},
    volume = {9},
    url = {https://doi.org/10.1038/s41534-023-00730-8}
}

@article{Anurag_jcc2026,
    author = {Anurag, K. S. V. and Patra, A. K. and Ghevade, V. D. and Sai Shankar, P. and Bhat, R. and Raghavendra, V. and Maitra, R. and Jaiganesh, G.},
    title = {{Resource Estimation for VQE on Small Molecules: Impact of Fermion Mappings and Hamiltonian Reductions}},
    journal = {Journal of Computational Chemistry},
    year = {2026},
    volume = {47},
    url = {https://doi.org/10.1002/jcc.70379}
}

@article{Cesario_pra2025,
    author = {Cesário, A. T. and Debarba, T.},
    title = {{Thermodynamic constraints on information transmission in quantum ensembles}},
    journal = {Physical Review A},
    year = {2025},
    volume = {112},
    url = {https://doi.org/10.1103/vjvj-sz12}
}

@article{Yamamoto_prr2022,
  title = {{Quantum hardware calculations of periodic systems with partition-measurement symmetry verification: Simplified models of hydrogen chain and iron crystals}},
  journal = {Physical Review Research},
  author = {Yamamoto, K. and Manrique, D. Z. and Khan, I. T. and Sawada, H. and Muñoz Ramo, D.},
  year = {2022},
  volume = {4},
  url = {https://doi.org/10.1103/PhysRevResearch.4.033110},
}

@article{Sherbert_rsc2021,
  title = {{A systematic variational approach to band theory in a quantum computer}},
  journal = {RSC Advances},
  author = {Sherbert, K. and Cerasoli, F. and Buongiorno Nardelli, M.},
  year = {2021},
  volume = {11},
  url = {https://doi.org/10.1039/d1ra07451b},
}

@article{Sherbert_sr2022,
  title = {{Quantum algorithm for electronic band structures with local tight‑binding orbitals}},
  journal = {Scientific Reports},
  author = {Sherbert, K. and Jayaraj, A. and Buongiorno Nardelli, M.},
  year = {2022},
  volume = {12},
  url = {https://doi.org/10.1038/s41598-022-13627-x},
}

@article{Khandelwal_sr2025,
  title = {{Quantum computation of the electronic structure of some prototype solids}},
  journal = {Scientific Reports},
  author = {Khandelwal, N. and Verma, N. and Jamdagni, P. and Kumar, A.},
  year = {2025},
  volume = {15},
  url = {https://doi.org/10.1038/s41598-025-27675-6},
}

@article{Dong_jpcc2023,
  title = {{Strongly Correlated Electronic Properties of FeO Studied by the SCAN+U Functional}},
  journal = {The Journal of Physical Chemistry C},
  author = {Dong, J. and Yan, X.-W. and Ma, F.},
  year = {2023},
  volume = {127},
  url = {https://doi.org/10.1021/acs.jpcc.2c07961},
}

@article{Mitas_prl2008,
  title = {{Quantum Monte Carlo Calculations of Structural Properties of FeO Under Pressure}},
  journal = {Physical Review Letters},
  author = {Kolorenč, J. and Mitas, L.},
  year = {2008},
  volume = {101},
  url = {https://doi.org/10.1103/PhysRevLett.101.185502},
}

@article{Mazin_am1998,
  title = {{Possible polytypism in FeO at high pressures}},
  journal = {American Mineralogist},
  author = {Mazin, I. I. and Fei, Y. and Downs, R. and Cohen, R.},
  year = {1998},
  volume = {83},
  url = {https://doi.org/10.2138/am-1998-5-604},
}

@article{Fang_prl1998,
  title = {{Inverse versus Normal NiAs Structures as High-Pressure Phases of FeO and MnO}},
  journal = {Physical Review Letters},
  author = {Fang, Z. and Terakura, K. and Sawada, H. and Miyazaki, T. and Solovyev, I.},
  year = {1998},
  volume = {81},
  url = {https://doi.org/10.1103/PhysRevLett.81.1027},
}

@article{Fang_prb1999,
  title = {{First-principles study on electronic structures and phase stability of MnO and FeO under high pressure}},
  journal = {Physical Review B.},
  author = {Fang, Z. and Solovyev, I. V. and Sawada, H. and Terakura, K.},
  year = {1999},
  volume = {59},
  url = {https://doi.org/10.1103/PhysRevB.59.762},
}

@article{Abdallah_prb2024,
    title = {{Quasiparticle self-consistent GW with effective vertex corrections in the polarizability and the self-energy applied to MnO, FeO, CoO, and NiO}},
    journal = {Physical Review B.},
    author = {Abdallah, M. S. and Pasquarello, A.},
    year = {2024},
    volume = {110},
    url = {https://doi.org/10.1103/PhysRevB.110.155105},
}

@article{Ho_natcomm2024,
    title = {{Quantum critical phase of FeO spans conditions of Earth's lower mantle}},
    journal = {Nature Communications.},
    author = {Ho, W.-G. D. and Zhang, P. and Haule, K. and Jackson, J. M. and Dobrosavljević, V. and Dobrosavljevic, V. V.},
    year = {2024},
    volume = {15},
    url = {https://doi.org/10.1038/s41467-024-47489-w},
}

@article{Leonov_prb2015,
    title = {{Metal-insulator transition and local-moment collapse in FeO under pressure}},
    journal = {Physical Review B.},
    author = {Leonov, I.},
    year = {2015},
    volume = {92},
    url = {https://doi.org/10.1103/PhysRevB.92.085142},
}

@article{Leonov_prb2020,
    title = {{Emergence of quantum critical charge and spin-state fluctuations near the pressure-induced Mott transition in MnO, FeO, CoO, and NiO}},
    journal = {Physical Review B.},
    author = {Leonov, I. and Shorikov, A. O. and Anisimov, V. I. and Abrikosov, I. A.},
    year = {2020},
    volume = {101},
    url = {https://doi.org/10.1103/PhysRevB.101.245144},
}

@article{Rodl_prb2012,
    title = {{Optical and energy-loss spectra of the antiferromagnetic transition metal oxides MnO, FeO, CoO, and NiO including quasiparticle and excitonic effects}},
    journal = {Physical Review B.},
    author = {Rödl, C. and Bechstedt, F.},
    year = {2012},
    volume = {86},
    url = {https://doi.org/10.1103/PhysRevB.86.235122},
}

@article{Rodl_prb2009,
    title = {{Quasiparticle band structures of the antiferromagnetic transition-metal oxides MnO, FeO, CoO, and NiO}},
    journal = {Physical Review B.},
    author = {Rödl, C. and Fuchs, F. and Furthmüller, J. and Bechstedt, F.},
    year = {2009},
    volume = {79},
    url = {https://doi.org/10.1103/PhysRevB.79.235114},
}

@article{Shorikov_prb2010,
    title = {{Orbital-selective pressure-driven metal to insulator transition in FeO from dynamical mean-field theory}},
    journal = {Physical Review B.},
    author = {Shorikov, A. O. and Pchelkina, Z. V. and Anisimov, V. I. and Skornyakov, S. L. and Korotin, M. A.},
    year = {2010},
    volume = {82},
    url = {https://doi.org/10.1103/PhysRevB.82.195101},
}

@article{Selisko_npjcompmat2025,
    title = {{Dynamical mean field theory for real materials on a quantum computer}},
    journal = {npj Computational Materials.},
    author = {Selisko, J. and Amsler, M. and Wever, C. and Kawashima, Y. and Samsonidze, G. and Ul Haq, R. and Tacchino, F. and Tavernelli, I. and Eckl, T.},
    year = {2025},
    volume = {11},
    url = {https://doi.org/10.1038/s41524-025-01772-6},
}

@article{Lee_arxiv2026,
    title = {{Benchmarking quantum simulation with neutron-scattering experiments}},
    journal = {arXiv.},
    author = {Lee, Y.-T. and Kumaran, K. and Pokharel, B. and Scheie, A. and Sarkis, C. L. and Tennant, D. A. and Humble, T. and Schleife, A. and Kandala, A. and Banerjee, A.},
    year = {2026},
    volume = {arXiv:2603.15608},
    url = {https://doi.org/10.48550/arXiv.2603.15608},
}

@article{Cao_faraday1998,
    title = {{Low-lying electronic states and molecular structure of Fe$_2$O$_2$}},
    journal = {Journal of the Chemical Society, Faraday Transactions},
    author = {Cao, Z. and Duran, M. and Solà, M.},
    year = {1998},
    volume = {94},
    url = {https://doi.org/10.1039/a803725f},
}

@article {Giannozzi_jpcm2009,
	author = {Giannozzi, P. and Baroni, S. and Bonini, N. and Calandra, M. and Car, R. and Cavazzoni, C. and Ceresoli, D. and Chiarotti, G. L. and Cococcioni, M. and Dabo, I. and Dal Corso, A. and de Gironcoli, S. and others},
	title = {{QUANTUM ESPRESSO: a modular and open-source software project for quantum simulations of materials}},
	volume = {21},
	year = {2009},
	journal = {Journal of Physics: Condensed Matter},
	url = {https://doi.org/10.1088/0953-8984/21/39/395502},
}

@article {Giannozzi_jpcm2017,
	author = {Giannozzi, P. and Andreussi, O. and Brumme, T. and Bunau, O. and Buongiorno Nardelli, M. and Calandra, M. and Car, R. and Cavazzoni, C. and Ceresoli, D. and Cococcioni, M. and Colonna, N. and others},
	title = {{Advanced capabilities for materials modelling with Quantum ESPRESSO}},
	volume = {29},
	year = {2017},
	journal = {Journal of Physics: Condensed Matter},
	url = {https://doi.org/10.1088/1361-648X/aa8f79},
}

@article{Perdew_prl1996,
  title = {Generalized gradient approximation made simple},
  author = {Perdew, J. P. and Burke, K. and Ernzerhof, M.},
  year = {1996},
  journal = {Physical Review Letters},
  volume = {77},
  pages = {3865},
  doi = {10.1103/PhysRevLett.77.3865},
  url = {https://doi.org/10.1103/PhysRevLett.77.3865}
}

@article{kincaidCorrelationConsistentEffective2022,
  title = {Correlation Consistent Effective Core Potentials for Late 3d Transition Metals Adapted for Plane Wave Calculations},
  author = {Kincaid, Benjamin and Wang, Guangming and Zhou, Haihan and Mitas, Lubos},
  year = {2022},
  month = nov,
  journal = {The Journal of Chemical Physics},
  volume = {157},
  number = {17},
  pages = {174307},
  publisher = {American Institute of Physics},
  issn = {0021-9606},
  doi = {10.1063/5.0109098},
  urldate = {2022-11-14}
}

@article{wang_new_2022,
  title = {A New Generation of Effective Core Potentials from Correlated and Spin--Orbit Calculations: {{Selected}} Heavy Elements},
  shorttitle = {A New Generation of Effective Core Potentials from Correlated and Spin--Orbit Calculations},
  author = {Wang, Guangming and Kincaid, Benjamin and Zhou, Haihan and Annaberdiyev, Abdulgani and Bennett, M. Chandler and Krogel, Jaron T. and Mitas, Lubos},
  year = {2022},
  month = aug,
  journal = {The Journal of Chemical Physics},
  volume = {157},
  number = {5},
  pages = {054101},
  publisher = {American Institute of Physics},
  issn = {0021-9606},
  doi = {10.1063/5.0087300},
  urldate = {2022-08-02}
}

@article {Mostofi_cpc2014,
	author = {Mostofi, A. A. and Yates, J. R. and Pizzi, G. and Lee, Y.-S. and Souza, I. and Vanderbilt, D. and Marzari, N.},
	title = {{An updated version of Wannier90: A tool for obtaining maximally-localised Wannier functions}},
	volume = {},
	year = {2014},
	journal = {Computer Physics Communications},
	url = {https://doi.org/10.1016/j.cpc.2014.05.003},
}

@article {Sakellaris_jcp2011,
	author = {Sakellaris, C. N. and Miliordos, E. and Mavridis, A.},
	title = {{First principles study of the ground and excited states of FeO, FeO+, and FeO-}},
	volume = {134},
	year = {2011},
	journal = {The Journal of Chemical Physics},
	url = {https://doi.org/10.1063/1.3598529},
}

@article {Cheung_jms1982,
	author = {Cheung, A. S.-C. and Lee, N. and Lyyra, A. M. and Merer, A. J. and Taylor, A. W.},
	title = {{Spectroscopic properties of the $^5 \Delta_i$ ground state of FeO}},
	volume = {95},
	year = {1982},
	journal = {Journal of Molecular Spectroscopy},
	url = {https://doi.org/10.1016/0022-2852(82)90248-X},
}

@article {Drechsler_jcp1997,
	author = {Drechsler, G. and Boesl, U. and Bäßmann, C. and Schlag, E. W.},
	title = {{Mass selected anion-zero kinetic energy photoelectron spectroscopy (anion-ZEKE): Ground and low excited states of FeO}},
	volume = {107},
	year = {1997},
	journal = {The Journal of Chemical Physics},
	url = {https://doi.org/10.1063/1.474622},
}

@article{Kulik_jcp2011,
  title = {{Accurate potential energy surfaces with a DFT+U(R) approach}},
  journal = {The Journal of Chemical Physics},
  author = {Kulik, H. and Marzari, N.},
  year = {2011},
  volume = {135},
  url = {https://doi.org/10.1063/1.3660353},
}

@article{Shee_rsc2023,
  title = {{Bridging physical intuition and hardware efficiency for correlated electronic states: the local unitary cluster Jastrow ansatz for electronic structure}},
  journal = {Chemical Science},
  author = {Motta, M. and Sung, K. J. and Whaley, K. B. and Head-Gordon, M. and Shee, J.},
  year = {2023},
  volume = {14},
  url = {https://doi.org/10.1039/d3sc02516k},
}

@article{Hamann_prb2013,
  title     = {Optimized norm-conserving Vanderbilt pseudopotentials},
  author    = {Hamann, D. R.},
  journal   = {Physical Review B},
  volume    = {88},
  issue     = {8},
  pages     = {085117},
  year      = {2013},
  publisher = {American Physical Society},
  doi       = {10.1103/PhysRevB.88.085117},
  url       = {https://link.aps.org/doi/10.1103/PhysRevB.88.085117}
}

@article{heyraud2023efficient,
  title={Efficient estimation of trainability for variational quantum circuits},
  author={Heyraud, Valentin and Li, Zejian and Donatella, Kaelan and Boit{\'e}, Alexandre Le and Ciuti, Cristiano},
  journal={PRX Quantum},
  volume={4},
  number={4},
  pages={040335},
  year={2023},
  publisher={APS},
  doi={10.1103/PRXQuantum.4.040335},
  url={https://journals.aps.org/prxquantum/abstract/10.1103/PRXQuantum.4.040335}
}

@article{anschuetz2021critical,
title={Critical Points in Quantum Generative Models},
author={Eric R. Anschuetz},
journal={International Conference on Learning Representations},
year={2022},
url={https://openreview.net/forum?id=2f1z55GVQN}
}

@article{robledo2025chemistry,
  title={Chemistry beyond the scale of exact diagonalization on a quantum-centric supercomputer},
  author={Robledo-Moreno, Javier and Motta, Mario and Haas, Holger and Javadi-Abhari, Ali and Jurcevic, Petar and Kirby, William and Martiel, Simon and Sharma, Kunal and Sharma, Sandeep and Shirakawa, Tomonori and others},
  journal={Science Advances},
  volume={11},
  number={25},
  pages={eadu9991},
  year={2025},
  publisher={American Association for the Advancement of Science},
url={https://www.science.org/doi/10.1126/sciadv.adu9991},
doi={10.1126/sciadv.adu9991}
}

@article{wang2023trainability,
    title={Trainability enhancement of parameterized quantum circuits via reduced-domain parameter initialization},
  author={Wang, Yabo and Qi, Bo and Ferrie, Chris and Dong, Daoyi},
  journal={Physical Review Applied},
  volume={22},
  number={5},
  pages={054005},
  year={2024},
  publisher={APS},
url={https://journals.aps.org/prapplied/abstract/10.1103/PhysRevApplied.22.054005},
doi={10.1103/PhysRevApplied.22.054005}
}

@article{mcclean2018barren,
  title={Barren plateaus in quantum neural network training landscapes},
  author={McClean, Jarrod R and Boixo, Sergio and Smelyanskiy, Vadim N and Babbush, Ryan and Neven, Hartmut},
  journal={Nature {C}ommunications},
  volume={9},
  number={1},
  pages={1--6},
  year={2018},
  publisher={Nature Publishing Group},
  url={https://doi.org/10.1038/s41467-018-07090-4},
  doi={10.1038/s41467-018-07090-4}
}

@article{peruzzo2014variational,
  title={A variational eigenvalue solver on a photonic quantum processor},
  author={Peruzzo, Alberto and McClean, Jarrod and Shadbolt, Peter and Yung, Man-Hong and Zhou, Xiao-Qi and Love, Peter J and Aspuru-Guzik, Al{\'a}n and O’brien, Jeremy L},
  journal={Nature {C}ommunications},
  volume={5},
  number={1},
  pages={1--7},
  year={2014},
  publisher={Nature Publishing Group},
  doi={10.1038/ncomms5213},
  url={https://www.nature.com/articles/ncomms5213#citeas}
}

@article{cerezo2020variationalreview,
   title={Variational quantum algorithms},
   author={Cerezo, M. and Arrasmith, Andrew and Babbush, Ryan and Benjamin, Simon C and Endo, Suguru and Fujii, Keisuke and McClean, Jarrod R and Mitarai, Kosuke and Yuan, Xiao and Cincio, Lukasz and Coles, Patrick J. },
   journal={Nature Reviews Physics},
   volume={3},
   number={1},
   pages={625–644},
   publisher={Nature Publishing Group},
   year={2021},
   url={https://www.nature.com/articles/s42254-021-00348-9},
   doi={10.1038/s42254-021-00348-9}
 }

@article{fontana2022nontrivial,
  title={Non-trivial symmetries in quantum landscapes and their resilience to quantum noise},
  author={Fontana, Enrico and Cerezo, M and Arrasmith, Andrew and Rungger, Ivan and Coles, Patrick J},
  journal={Quantum},
  volume={6},
  pages={804},
  year={2022},
  publisher={Verein zur F{\"o}rderung des Open Access Publizierens in den Quantenwissenschaften},
  url={https://quantum-journal.org/papers/q-2022-09-15-804/},
  doi={10.22331/q-2022-09-15-804}
}

@article{larocca2021diagnosing,
  doi = {10.22331/q-2022-09-29-824},
  url = {https://doi.org/10.22331/q-2022-09-29-824},
  title = {Diagnosing {B}arren {P}lateaus with {T}ools from {Q}uantum {O}ptimal {C}ontrol},
  author = {Larocca, Martin and Czarnik, Piotr and Sharma, Kunal and Muraleedharan, Gopikrishnan and Coles, Patrick J. and Cerezo, M.},
  journal = {{Quantum}},
  issn = {2521-327X},
  publisher = {{Verein zur F{\"{o}}rderung des Open Access Publizierens in den Quantenwissenschaften}},
  volume = {6},
  pages = {824},
  month = sep,
  year = {2022}
}

@article{larocca2024review,
  title={Barren plateaus in variational quantum computing},
  author={Larocca, Martin and Thanasilp, Supanut and Wang, Samson and Sharma, Kunal and Biamonte, Jacob and Coles, Patrick J and Cincio, Lukasz and McClean, Jarrod R and Holmes, Zo{\"e} and Cerezo, M},
   journal={Nature Reviews Physics},
   volume={7},
   number={4},
   pages={174–189},
   publisher={Nature Publishing Group},
   year={2025},
   url={https://www.nature.com/articles/s42254-025-00813-9},
   doi={10.1038/s42254-025-00813-9}
 }

@article{cerezo2020cost,
  title={Cost function dependent barren plateaus in shallow parametrized quantum circuits},
  author={Cerezo, M. and Sone, Akira and Volkoff, Tyler and Cincio, Lukasz and Coles, Patrick J},
  journal={Nature {C}ommunications},
  volume={12},
  number={1},
  pages={1--12},
  year={2021},
  publisher={Nature Publishing Group},
  url = {https://doi.org/10.1038/s41467-021-21728-w},
  doi={10.1038/s41467-021-21728-w}
}

@inproceedings{zhang2022escaping,
  title={Escaping from the Barren Plateau via {G}aussian Initializations in Deep Variational Quantum Circuits},
  author={Zhang, Kaining and Liu, Liu and Hsieh, Min-Hsiu and Tao, Dacheng},
  booktitle={Advances in Neural Information Processing Systems},
  url={https://openreview.net/forum?id=jXgbJdQ2YIy},
  year={2022}
}

@article{park2023hamiltonian,
  title={Hamiltonian variational ansatz without barren plateaus},
  author={Park, Chae-Yeun and Killoran, Nathan},
  journal={Quantum},
  volume={8},
  pages={1239},
  year={2024},
  publisher={Verein zur F{\"o}rderung des Open Access Publizierens in den Quantenwissenschaften},
  url={https://quantum-journal.org/papers/q-2024-02-01-1239/},
  doi={10.22331/q-2024-02-01-1239}
}

@article{shi2024avoiding,
  title={Avoiding barren plateaus via {G}aussian Mixture Model},
  author={Shi, Xiao and Shang, Yun},
  journal={arXiv preprint arXiv:2402.13501},
  year={2024},
url={https://arxiv.org/abs/2402.13501}
}

@article{park2024hardware,
  title={Hardware-efficient ansatz without barren plateaus in any depth},
  author={Park, Chae-Yeun and Kang, Minhyeok and Huh, Joonsuk},
  journal={arXiv preprint arXiv:2403.04844},
  year={2024},
url={https://arxiv.org/abs/2403.04844}
}

@article{holmes2021connecting,
 title={Connecting ansatz expressibility to gradient magnitudes and barren plateaus},
 author={Holmes, Zo{\"e} and Sharma, Kunal and Cerezo, M. and Coles, Patrick J},
journal={PRX Quantum},
 year={2022},
volume = {3},
  issue = {1},
  pages = {010313},
  month = {Jan},
  publisher = {American Physical Society},
  doi = {10.1103/PRXQuantum.3.010313},
  url = {https://doi.org/10.1103/PRXQuantum.3.010313}
}

@article{grant2019initialization,
  title={An initialization strategy for addressing barren plateaus in parametrized quantum circuits},
  author={Grant, Edward and Wossnig, Leonard and Ostaszewski, Mateusz and Benedetti, Marcello},
  journal={Quantum},
  volume={3},
  pages={214},
  year={2019},
  publisher={Verein zur F{\"o}rderung des Open Access Publizierens in den Quantenwissenschaften},
  url={https://quantum-journal.org/papers/q-2019-12-09-214/},
  doi={10.22331/q-2019-12-09-214}
}

@article{duffield2023bayesian,
  title={Bayesian learning of parameterised quantum circuits},
  author={Duffield, Samuel and Benedetti, Marcello and Rosenkranz, Matthias},
  journal={Machine Learning: Science and Technology},
  volume={4},
  number={2},
  pages={025007},
  year={2023},
  publisher={IOP Publishing},
  url={https://iopscience.iop.org/article/10.1088/2632-2153/acc8b7},
  doi={10.1088/2632-2153/acc8b7}
}

@inproceedings{kulshrestha2022beinit,
  title={Beinit: Avoiding barren plateaus in variational quantum algorithms},
  author={Kulshrestha, Ankit and Safro, Ilya},
  booktitle={2022 IEEE International Conference on Quantum Computing and Engineering (QCE)},
  pages={197--203},
  year={2022},
  organization={IEEE},
  url={https://ieeexplore.ieee.org/stamp/stamp.jsp?arnumber=9951195},
  doi={10.1109/QCE53715.2022.00039}
}

@inproceedings{you2021exponentially,
  title={Exponentially many local minima in quantum neural networks},
  author={You, Xuchen and Wu, Xiaodi},
  booktitle={International Conference on Machine Learning},
  pages={12144--12155},
  year={2021},
  organization={PMLR},
  url={https://proceedings.mlr.press/v139/you21c.html}
}

@article{anschuetz2022beyond,
  title={Beyond Barren Plateaus: Quantum Variational Algorithms Are Swamped With Traps},
  author={Anschuetz, Eric R and Kiani, Bobak T},
  journal={Nature Communications},
  volume={13},
  number={1},
  pages={7760},
  year={2022},
  publisher={Nature Publishing Group UK London},
  doi={10.1038/s41467-022-35364-5},
  url={https://doi.org/10.1038/s41467-022-35364-5},
}

@misc{qiskit,
    author = {{Qiskit contributors}},
    title = {Qiskit: An Open-source Framework for Quantum Computing},
    year = {2023},
    doi = {10.5281/zenodo.2573505}
}

@article{puig2024variational,
  title = {Variational Quantum Simulation: A Case Study for Understanding Warm Starts},
  author = {Puig, Ricard and Drudis, Marc and Thanasilp, Supanut and Holmes, Zo\"e},
  journal = {PRX Quantum},
  volume = {6},
  issue = {1},
  pages = {010317},
  numpages = {39},
  year = {2025},
  month = {Jan},
  publisher = {American Physical Society},
  doi = {10.1103/PRXQuantum.6.010317},
  url = {https://link.aps.org/doi/10.1103/PRXQuantum.6.010317}
}

@article{kanno2023quantum,
  title={Quantum-selected configuration interaction: Classical diagonalization of Hamiltonians in subspaces selected by quantum computers},
  author={Kanno, Keita and Kohda, Masaya and Imai, Ryosuke and Koh, Sho and Mitarai, Kosuke and Mizukami, Wataru and Nakagawa, Yuya O},
  journal={arXiv preprint arXiv:2302.11320},
  year={2023},
  url={https://arxiv.org/abs/2302.11320},
  doi={10.48550/arXiv.2302.11320}
}

@article{kulshrestha2026exponentially,
  title={Exponentially many initializations to avoid barren plateaus},
  author={Kulshrestha, Ankit and Puig, Ricard and Garc{\'\i}a-Mart{\'\i}n, Diego and Cincio, Lukasz and Safro, Ilya and Holmes, Zo{\"e} and Cerezo, M},
  journal={arXiv preprint arXiv:2606.18515},
  year={2026},
  doi = {https://doi.org/10.48550/arXiv.2606.18515},
  url = {https://arxiv.org/abs/2606.18515}
}

@article{lin2025improved,
  title={Improved parameter initialization for the (local) unitary cluster Jastrow ansatz},
  author={Lin, Wan-Hsuan and Liang, Fangchun and Motta, Mario and Zhang, Haimeng and Merz Jr, Kenneth M and Sung, Kevin J},
  journal={arXiv preprint arXiv:2511.22476},
  year={2025},
  url={https://arxiv.org/abs/2511.22476},
  doi={10.48550/arXiv.2511.22476
}
}

%\makeatletter\@input{xx.tex}\makeatother
%\end{document}

% ================= SUPPLEMENT =================
\clearpage
\appendix
\section*{\label{sec:supp_inf}Supplementary Information for: ``Towards Quantum Computation of Bulk FeO''}

\appendix
\section{Density functional theory}
\label{sec:si_dft_images}
The simulation cell size for the Density Functional Theory (DFT) calculations performed in the main manuscript was determined via band structure calculations. If the simulation cell is sufficiently large, the periodic images do not interact with each other and the bands should be completely flat. The non-polarized PBE band structure for the FeO molecule in a 15$\times$15$\times$15 Å$^3$ simulation cell is shown in Figure \ref{fig:si_feo_bands}. 

\setcounter{figure}{0} % Resets the figure counter to 0 so the next figure becomes 1
\renewcommand{\thefigure}{S\arabic{figure}} % Prepends 'S' to the figure number
\begin{figure}[!h]
\centering
\includegraphics[width=0.48\textwidth]{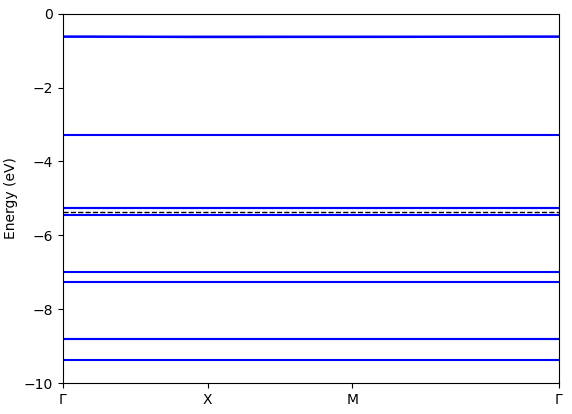}
\caption{FeO band structure calculated via non-spin-polarized Density Functional Theory within the PBE approximation. The dotted line denotes the Fermi energy.
}
\label{fig:si_feo_bands}
\end{figure}

This band structure calculation was performed using Optimized Norm-Conserving Vanderbilt Pseudopotentials (ONCV)\cite{kincaidCorrelationConsistentEffective2022, wang_new_2022} rather than the correlation-consistent Effective Core Potentials used in the main manuscript,\cite{Hamann_prb2013} but this choice does not affect the magnitude of inter-image hops. The plane wave cutoff was also slightly lower (240 Ry) than the value employed in the main manuscript (320 Ry), but this is more than sufficient for the Fe and O ONCV pseudopotentials.

\section{Hopping amplitudes}
\label{sec:si_hopping}
The nonzero hopping amplitudes in FeO as calculated via Wannier90 are displayed in Table \ref{tab:hops_feo}. Note that the hopping between the $d_{z^2}$ and $d_{x^2-y^2}$ is a consequence of the crystal field and hybridization included in the DFT from which the hopping amplitudes are determined. The $t_{ij}$ denote hopping between distinct orbitals, while $\varepsilon_j$ indicates ``hopping" from and to the same Wannier function, which corresponds to the onsite energy of that Wannier function. The onsite energies and nonzero hopping parameters for the Fe$_2$O$_2$ molecule are tabulated in Table \ref{tab:hops_fe2o2} where the superscript indexes the two possible choices of Fe and O atom.

\setcounter{table}{0}
\renewcommand{\thetable}{S\arabic{table}} 
\begin{table}[h]
    \caption{Onsite energies and nonzero hopping amplitudes in the FeO molecule calculated via DFT/Wannier90}
    \label{tab:hops_feo}
    \centering
    \begin{tabular}{cccc}
        \hline\hline
        $w_i$ & $w_j$ & $t_{ij}$ (eV) & $\varepsilon_i$ (eV) \\
        \hline
        $d_{xy}$ & $d_{xy}$ & - & -6.0 \\
        $d_{xy}$ & $p_y$ & -1.8 & - \\
        $d_{yz}$ & $d_{yz}$ & - & -4.9 \\
        $d_{z^2}$ & $d_{z^2}$ & - & -5.3 \\
        $d_{z^2}$ & $d_{x^2-y^2}$ & 0.7 & - \\
        $d_{z^2}$ & $p_x$ & -1.0 & - \\
        $d_{xz}$ & $d_{xz}$ & - & -6.1 \\
        $d_{xz}$ & $p_z$ & -1.8 & - \\
        $d_{x^2-y^2}$ & $d_{x^2-y^2}$ & - & -6.0 \\
        $d_{x^2-y^2}$ & $p_x$ & 1.8 & - \\
        $p_x$ & $p_x$ & - & -8.1 \\
        $p_y$ & $p_y$ & - & -7.5 \\
        $p_z$ & $p_z$ & - & -7.4 \\
        \hline\hline
    \end{tabular}
\end{table}

\begin{table}[h]
    \caption{Onsite energies and nonzero hopping amplitudes in the Fe$_2$O$_2$ molecule calculated via DFT/Wannier90. Hops with magnitude less than 0.1 eV were set to 0 in our model.}
    \label{tab:hops_fe2o2}
    \centering
    \small
    \renewcommand{\arraystretch}{1.35}
    \begin{tabular}{cccc}
        \hline\hline
        $w_i$ & $w_j$ & $t_{ij}$ (eV) & $\varepsilon_i$ (eV) \\
        \hline

        $d_{z^2}^{\mathrm{Fe}_{1,2}}$     & $d_{z^2}^{\mathrm{Fe}_{1,2}}$     & -     & -4.64 \\
        $d_{xz}^{\mathrm{Fe}_{1,2}}$      & $d_{xz}^{\mathrm{Fe}_{1,2}}$      & -     & -4.69 \\
        $d_{yz}^{\mathrm{Fe}_{1,2}}$      & $d_{yz}^{\mathrm{Fe}_{1,2}}$      & -     & -4.67 \\
        $d_{x^2-y^2}^{\mathrm{Fe}_{1,2}}$ & $d_{x^2-y^2}^{\mathrm{Fe}_{1,2}}$ & -     & -5.16 \\
        $d_{xy}^{\mathrm{Fe}_{1,2}}$      & $d_{xy}^{\mathrm{Fe}_{1,2}}$      & -     & -4.55 \\
        $p_z^{\mathrm{O}_{1,2}}$          & $p_z^{\mathrm{O}_{1,2}}$          & -     & -7.347 \\
        $p_x^{\mathrm{O}_{1,2}}$          & $p_x^{\mathrm{O}_{1,2}}$          & -     & -8.284 \\
        $p_y^{\mathrm{O}_{1,2}}$          & $p_y^{\mathrm{O}_{1,2}}$          & -     & -7.452 \\

        \hline

        $d_{z^2}^{\mathrm{Fe}_1}$     & $d_{x^2-y^2}^{\mathrm{Fe}_1}$ & -0.08  & - \\
        $d_{z^2}^{\mathrm{Fe}_2}$     & $d_{x^2-y^2}^{\mathrm{Fe}_2}$ & -0.08  & - \\

        $p_z^{\mathrm{O}_1}$          & $p_z^{\mathrm{O}_2}$          & -0.432 & - \\
        $p_x^{\mathrm{O}_1}$          & $p_x^{\mathrm{O}_2}$          & -0.394 & - \\
        $p_y^{\mathrm{O}_1}$          & $p_y^{\mathrm{O}_2}$          &  0.882 & - \\

        \hline

        $d_{z^2}^{\mathrm{Fe}_1}$     & $p_x^{\mathrm{O}_1}$          & -0.7 & - \\
        $d_{z^2}^{\mathrm{Fe}_1}$     & $p_y^{\mathrm{O}_1}$          &  0.5 & - \\
        $d_{xz}^{\mathrm{Fe}_1}$      & $p_z^{\mathrm{O}_1}$          & -1.1 & - \\
        $d_{yz}^{\mathrm{Fe}_1}$      & $p_z^{\mathrm{O}_1}$          &  1.1 & - \\
        $d_{x^2-y^2}^{\mathrm{Fe}_1}$ & $p_x^{\mathrm{O}_1}$          & -0.8 & - \\
        $d_{x^2-y^2}^{\mathrm{Fe}_1}$ & $p_y^{\mathrm{O}_1}$          & -1.1 & - \\
        $d_{xy}^{\mathrm{Fe}_1}$      & $p_x^{\mathrm{O}_1}$          & -1.4 & - \\
        $d_{xy}^{\mathrm{Fe}_1}$      & $p_y^{\mathrm{O}_1}$          &  1.4 & - \\

        $d_{z^2}^{\mathrm{Fe}_1}$     & $p_x^{\mathrm{O}_2}$          & -0.7 & - \\
        $d_{z^2}^{\mathrm{Fe}_1}$     & $p_y^{\mathrm{O}_2}$          & -0.5 & - \\
        $d_{xz}^{\mathrm{Fe}_1}$      & $p_z^{\mathrm{O}_2}$          & -1.1 & - \\
        $d_{yz}^{\mathrm{Fe}_1}$      & $p_z^{\mathrm{O}_2}$          & -1.1 & - \\
        $d_{x^2-y^2}^{\mathrm{Fe}_1}$ & $p_x^{\mathrm{O}_2}$          & -0.8 & - \\
        $d_{x^2-y^2}^{\mathrm{Fe}_1}$ & $p_y^{\mathrm{O}_2}$          &  1.1 & - \\
        $d_{xy}^{\mathrm{Fe}_1}$      & $p_x^{\mathrm{O}_2}$          &  1.4 & - \\
        $d_{xy}^{\mathrm{Fe}_1}$      & $p_y^{\mathrm{O}_2}$          &  1.4 & - \\

        $d_{z^2}^{\mathrm{Fe}_2}$     & $p_x^{\mathrm{O}_1}$          &  0.7 & - \\
        $d_{z^2}^{\mathrm{Fe}_2}$     & $p_y^{\mathrm{O}_1}$          &  0.5 & - \\
        $d_{xz}^{\mathrm{Fe}_2}$      & $p_z^{\mathrm{O}_1}$          &  1.1 & - \\
        $d_{yz}^{\mathrm{Fe}_2}$      & $p_z^{\mathrm{O}_1}$          &  1.1 & - \\
        $d_{x^2-y^2}^{\mathrm{Fe}_2}$ & $p_x^{\mathrm{O}_1}$          &  0.8 & - \\
        $d_{x^2-y^2}^{\mathrm{Fe}_2}$ & $p_y^{\mathrm{O}_1}$          & -1.1 & - \\
        $d_{xy}^{\mathrm{Fe}_2}$      & $p_x^{\mathrm{O}_1}$          & -1.4 & - \\
        $d_{xy}^{\mathrm{Fe}_2}$      & $p_y^{\mathrm{O}_1}$          & -1.4 & - \\

        $d_{z^2}^{\mathrm{Fe}_2}$     & $p_x^{\mathrm{O}_2}$          &  0.7 & - \\
        $d_{z^2}^{\mathrm{Fe}_2}$     & $p_y^{\mathrm{O}_2}$          & -0.5 & - \\
        $d_{xz}^{\mathrm{Fe}_2}$      & $p_z^{\mathrm{O}_2}$          &  1.1 & - \\
        $d_{yz}^{\mathrm{Fe}_2}$      & $p_z^{\mathrm{O}_2}$          & -1.1 & - \\
        $d_{x^2-y^2}^{\mathrm{Fe}_2}$ & $p_x^{\mathrm{O}_2}$          &  0.8 & - \\
        $d_{x^2-y^2}^{\mathrm{Fe}_2}$ & $p_y^{\mathrm{O}_2}$          &  1.1 & - \\
        $d_{xy}^{\mathrm{Fe}_2}$      & $p_x^{\mathrm{O}_2}$          &  1.4 & - \\
        $d_{xy}^{\mathrm{Fe}_2}$      & $p_y^{\mathrm{O}_2}$          & -1.4 & - \\

        \hline\hline
    \end{tabular}
\end{table}

\section{Variational circuit optimization}
\label{sec:si_vqe_training}

The FeO and Fe$_2$O$_2$ variational circuits are trained using the same
two-stage procedure. Since the systems considered here are sufficiently small
to admit exact classical evaluation within the relevant particle-number
sectors, the variational energies used during training are obtained using Qiskit~\cite{qiskit} 
noiseless statevector simulations. Thus, for a reference determinant $\ket{\Phi_r}$, the
objective function is evaluated exactly as
\begin{equation}
E_r(\boldsymbol{\theta})
=
\bra{\Psi_r(\boldsymbol{\theta})}
\hat H
\ket{\Psi_r(\boldsymbol{\theta})}.
\label{eq:si_vqe_cost}
\end{equation}
The quantum processor is subsequently used to sample occupation-number
configurations from the optimized circuits.

An important feature of the circuits used here is that they conserve
$N_\uparrow$ and $N_\downarrow$ independently. Each optimization therefore
takes place entirely within the $(N_\uparrow,N_\downarrow)$ sector of its
reference determinant. For FeO, separate reference determinants are used in
the sectors
\begin{equation}
(N_\uparrow,N_\downarrow)
=
(4,8),\,(5,7),\,(6,6),\,(6,6),\,(7,5),\,(8,4).
\end{equation}
The Hamiltonian is invariant under global spin reversal, so the
$(4,8)$ and $(8,4)$ calculations, as well as the $(5,7)$ and $(7,5)$
calculations, are related exactly. We therefore optimize only one member of
each spin-reversed pair and generate the partner circuit by exchanging the
spin-up and spin-down parameters. The two $(6,6)$ reference determinants are
similarly related by spin reversal. For Fe$_2$O$_2$, the calculations reported
in this work are performed in the
$(N_\uparrow,N_\downarrow)=(16,8)$ sector, and the variational circuit remains
in this sector throughout the optimization.

The optimization proceeds in two stages. We first optimize a
particle-number-preserving orbital-rotation ansatz,
\begin{equation}
\ket{\Psi_r^{\rm orb}(\boldsymbol{\kappa})}
=
\hat K(\boldsymbol{\kappa})
\ket{\Phi_r}.
\label{eq:si_orbital_ansatz}
\end{equation}
The operator $\hat K$ is a fermionic Gaussian unitary and therefore maps a
Slater determinant to another Slater determinant. This first optimization
determines the lowest-energy state accessible within the chosen
single-determinant manifold and provides a physically informed initialization
for the subsequent correlated calculation.

For each spin species, the orbital rotation is decomposed into a complete
brickwork network of nearest-neighbor Givens rotations. For $N_{\rm orb}$
spatial orbitals, a complete real orbital transformation contains
\begin{equation}
N_G
=
\frac{N_{\rm orb}(N_{\rm orb}-1)}{2}\,.
\end{equation}
Givens angles per spin register and hence $2N_G$ variational parameters in the
orbital-only circuit. Explicitly,
\begin{equation}
\hat K(\boldsymbol{\kappa})
=
\hat K_\uparrow(\boldsymbol{\kappa}_\uparrow)
\hat K_\downarrow(\boldsymbol{\kappa}_\downarrow),
\end{equation}
where each $\hat K_\sigma$ is implemented as a product of two-mode rotations
of the form
\begin{equation}
\hat G_{pq}(\theta)
=
\exp\left[
\theta
\left(
\hat a^\dagger_{p\sigma}\hat a_{q\sigma}
-
\hat a^\dagger_{q\sigma}\hat a_{p\sigma}
\right)
\right].
\end{equation}
For FeO, $N_{\rm orb}=8$, giving $N_G=28$ rotations per spin and $56$
orbital parameters in total. For Fe$_2$O$_2$, $N_{\rm orb}=16$, giving
$N_G=120$ rotations per spin and $240$ orbital parameters.

The optimized orbital circuit is subsequently embedded exactly into the
one-layer LUCJ ansatz
\begin{equation}
\ket{\Psi_r(\boldsymbol{\theta})}
=
\hat K_{\rm post}
\hat U_J
\hat K_{\rm pre}
\ket{\Phi_r}.
\label{eq:si_lucj_ansatz}
\end{equation}
Specifically, if
$\boldsymbol{\kappa}_r^\star$ denotes the parameters obtained from
Eq.~\eqref{eq:si_orbital_ansatz}, the initial LUCJ parameters are chosen such
that
\begin{equation}
\hat K_{\rm pre}
=
\hat K(\boldsymbol{\kappa}_r^\star),
\qquad
\hat U_J=\mathbb I,
\qquad
\hat K_{\rm post}=\mathbb I.
\label{eq:si_lucj_initialization}
\end{equation}
Consequently, the initial state of the LUCJ optimization is exactly the
optimized orbital-only state,
\begin{equation}
\ket{\Psi_r(\boldsymbol{\theta}_0)}
=
\ket{\Psi_r^{\rm orb}(\boldsymbol{\kappa}_r^\star)},
\end{equation}
and its initial energy is identical to the optimized orbital energy. This ensures that the LUCJ
optimization begins from a physically meaningful low-energy point rather than
from a random location in parameter space.

The Jastrow layer is
\begin{equation}
\hat U_J
=
\exp\left(
i\sum_{p=1}^{N_{\rm orb}}
J_p\,
\hat n_{p\uparrow}
\hat n_{p\downarrow}
\right),
\label{eq:si_local_jastrow}
\end{equation}
where
$\hat n_{p\sigma}=\hat a^\dagger_{p\sigma}\hat a_{p\sigma}$.
In the circuit representation, each term is implemented as a local
controlled-phase operation between the spin-up and spin-down modes associated
with the same spatial orbital. The locality of this interaction avoids the
all-to-all connectivity required by a fully general unitary cluster Jastrow
ansatz.

Both $\hat K_{\rm pre}$ and $\hat K_{\rm post}$ contain independent complete
orbital-rotation networks. The total number of parameters in the one-layer
LUCJ circuit is therefore
\begin{equation}
N_{\rm LUCJ}
=
4N_G+N_{\rm orb}.
\end{equation}
For FeO this gives
\begin{equation}
N_{\rm LUCJ}
=
2(56)+8
=
120,
\end{equation}
whereas for Fe$_2$O$_2$,
\begin{equation}
N_{\rm LUCJ}
=
2(240)+16
=
496.
\end{equation}

For both molecules, the orbital-only optimization is performed using the
derivative-free COBYLA algorithm. We use two initializations: one in which all
Givens angles are zero and a second weakly perturbed initialization with angles
drawn from a normal distribution of standard deviation $0.03$. Each orbital
optimization is allowed up to $800$ objective-function evaluations, with
initial COBYLA trust-region radius $0.08$. The lowest-energy result over the
two runs is retained.

The resulting optimized orbital parameters are then embedded into the LUCJ
circuit according to Eq.~\eqref{eq:si_lucj_initialization}. The full LUCJ
circuit is subsequently optimized with COBYLA for up to $1000$
objective-function evaluations, using an initial trust-region radius of
$0.04$. The optimization tolerance is $10^{-8}$. In this second stage we use
the embedded orbital solution itself as the initial point, so that any
improvement in energy relative to the orbital-only result is produced by the
additional variational freedom of the LUCJ ansatz. Throughout the optimization, the statevector is explicitly checked to remain
within the intended $(N_\uparrow,N_\downarrow)$ sector.

Finally, after the statevector optimization is completed, the optimized
parameter values are fixed and the corresponding circuits are compiled for
execution on quantum hardware. No hardware-in-the-loop optimization is
performed. The role of the quantum processor in the present workflow is
therefore to generate computational-basis samples from the trained
variational states, which are subsequently used to construct the determinant
spaces entering the SQD calculation.

\section{Hardware implementation of Sample-based Quantum Diagonalization}
\label{sec:si_hardware_sqd}

The SQD calculations use the optimized LUCJ circuits described in
Sec.~\ref{sec:si_vqe_training} as quantum samplers. The variational parameters
are first optimized in noiseless statevector simulations and are subsequently
held fixed during the hardware experiments. The parameter-bound circuits are
compiled to the IBM quantum processor \texttt{ibm\_miami} and measured in the
computational basis.

\subsection{Parallel circuit execution}

Because the individual molecular circuits occupy substantially fewer qubits
than are available on the processor, we increase the sampling throughput by
executing several circuits simultaneously on disjoint regions of the same
device. Suppose that $m$ nonoverlapping copies of a circuit are embedded on
the processor. We construct a single multiplexed circuit
\begin{equation}
\hat U_{\rm par}
=
\hat U^{(1)}
\otimes
\hat U^{(2)}
\otimes\cdots\otimes
\hat U^{(m)},
\end{equation}
where each $\hat U^{(j)}$ acts on a distinct set of physical qubits. The
subcircuits contain no gates coupling different regions and are executed in
parallel. Consequently, each hardware shot returns one molecular bitstring
from each region. Thus, $N_{\rm shot}$ executions of a circuit containing
$m$ spatially multiplexed instances produce $mN_{\rm shot}$ molecular
bitstrings, while requiring only $N_{\rm shot}$ QPU executions.

For FeO, each LUCJ circuit acts on 16 qubits. We embed six 16-qubit circuits
on disjoint regions of \texttt{ibm\_miami}, using 96 physical qubits in total.
The six circuits correspond to the six reference states discussed above and
are assigned separate quantum and classical registers. A single multiplexed
execution therefore returns six 16-bit occupation strings, one from each
reference circuit. The hardware run uses 20,000 shots, corresponding to
20,000 samples from each reference and 120,000 molecular bitstrings in total.

For Fe$_2$O$_2$, each circuit acts on 32 qubits. We similarly embed three
32-qubit copies on disjoint regions of the processor, again using 96 physical
qubits. Each QPU execution therefore produces three Fe$_2$O$_2$
occupation-number samples. In the hardware dataset used here, two
multiplexed runs of 450,000 shots are performed for each sampler, resulting
in six separate sample streams and $2.7\times10^6$ molecular samples per
sampler.

The physical layouts are fixed before compilation. For each region we verify
that all two-qubit couplings required by the LUCJ circuit are supported by
the current backend connectivity. Routing between regions is disabled, and
the compiled circuit is checked to contain neither SWAP gates nor two-qubit
operations connecting distinct molecular instances. The individual
measurement registers are retained separately so that the samples generated
by each physical realization can also be analyzed independently before
pooling.

\subsection{Hardware sampling and postselection}

The circuits are executed using the Qiskit Runtime Sampler. For the FeO
experiments we employ dynamical decoupling and gate and measurement twirling;
the 20,000-shot run is distributed over 20 twirling randomizations with
1,000 shots per randomization. The exact compiled circuit and the
corresponding hardware and compilation metadata are stored together with the
raw samples.

Next, we recall that while a noiseless  LUCJ circuit preserves $N_\uparrow$ and $N_\downarrow$
separately,  hardware errors can produce measured configurations
outside the target particle-number sector. Each sample stream is therefore
strictly postselected according to
\begin{equation}
N_\uparrow(x)=N_\uparrow^{\rm target},
\qquad
N_\downarrow(x)=N_\downarrow^{\rm target}.
\end{equation}
Samples that fail this condition are discarded. We do not apply
configuration recovery or modify measured bitstrings to restore the desired
particle number. 

The measured frequencies are used only to determine the selected determinant
support. Given the retained set $\mathcal S_0$, we construct the projected
Hamiltonian
\begin{equation}
H_{\mathcal S_0}
=
P_{\mathcal S_0} H P_{\mathcal S_0}
\end{equation}
\vspace{1.5mm}

\noindent
using the exact classical Hamiltonian matrix elements and diagonalize it
classically. We additionally enlarge the quantum-selected determinant space using the
connectivity of the Hamiltonian in Fock space. The postselected determinants
obtained from the quantum processor define the initial set
$\mathcal S_0$. We regard each determinant as a vertex of a graph and connect
two determinants $\ket{x}$ and $\ket{y}$ whenever the Hamiltonian has a
nonzero off-diagonal matrix element between them,
$\bra{y}H\ket{x}\neq0$. Starting from $\mathcal S_0$, we then define
\begin{equation}
\mathcal S_{m+1}
=
\mathcal S_m
\cup
\left\{
\ket{y}:
\exists\,\ket{x}\in\mathcal S_m,\,
\bra{y}H\ket{x}\neq0
\right\}.
\end{equation}
Thus, the one-shell space $\mathcal S_1$ contains the sampled determinants
together with every determinant directly coupled to them by the Hamiltonian.
The two-shell space $\mathcal S_2$ repeats the same construction once more,
and therefore contains all determinants that can be reached from the sampled
support through at most two nonzero Hamiltonian matrix elements.

For an electronic Hamiltonian, these edges correspond to the fermionic
processes appearing explicitly in $H$, such as orbital hopping and, when
present, interaction-induced transitions. Importantly, we determine the
connectivity directly from the nonzero matrix elements of the Hamiltonian
rather than imposing a separate excitation rule. The resulting enlarged
spaces can therefore contain determinants that were never observed on the
quantum computer, but which are directly coupled to configurations that were
sampled.

For each selected space we record its dimension, the lowest Ritz energy, and
the full-space residual. When studying convergence with shot number, smaller
datasets are formed from nested subsets of the measured samples so that the
different SQD calculations are compared using a common underlying hardware
dataset.

%\section*{References}
%\bibliography{supplement}

%\end{document}

\end{document}